\documentclass[a4paper]{article}

\usepackage{INTERSPEECH2019}

\title{Convolutional Neural Network-based Speech Enhancement for\\Cochlear Implant Recipients}
\name{Nursadul Mamun, Soheil Khorram, John H.L. Hansen}

\address{
  Cochlear Implant Processing Laboratory, Center for Robust Speech Systems (CRSS-CILab), 
Department of Electrical \& Computer Engineering, The University of Texas at Dallas
}
\email{(nursadul.mamun,soheil.khorram,john.hansen)@utdallas.edu}

\begin{document}
\maketitle
\begin{abstract}
Attempts to develop speech enhancement algorithms with improved speech intelligibility for cochlear implant (CI) users have met with limited success. To improve speech enhancement methods for CI users, we propose to perform speech enhancement in a cochlear filter-bank feature space, a feature-set specifically designed for CI users based on CI auditory stimuli. We leverage a convolutional neural network (CNN) to extract both stationary and non-stationary components of environmental acoustics and speech. We propose three CNN architectures: (1) vanilla CNN that directly generates the enhanced signal; (2) spectral-subtraction-style CNN (SS-CNN) that first predicts noise and then generates the enhanced signal by subtracting noise from the noisy signal; (3) Wiener-style CNN (Wiener-CNN) that generates an optimal mask for suppressing noise. An important problem of the proposed networks is that they introduce considerable delays, which limits their real-time application for CI users. To address this, this study also considers causal variations of these networks. Our experiments show that the proposed networks (both causal and non-causal forms) achieve significant improvement over existing baseline systems. We also found that causal Wiener-CNN outperforms other networks, and leads to the best overall envelope coefficient measure (ECM). The proposed algorithms represent a viable option for implementation on the CCi-MOBILE research platform as a pre-processor for CI users in naturalistic environments.
\end{abstract}

\section{Introduction}

 A cochlear implant (CI) is an implantable electronic device that provides the necessary sensation for hearing~\cite{lane2019cochlear, dieter2019near, ali2018cci}; CI partially restores hearing ability for subjects with sensorineural hearing loss (generally profound hearing loss). According to a report by the U.S. Food and Drug Administration, over 96000 people in US (324,000 people worldwide) have received a CI device by the end of 2012~\cite{NIDCD2014l}.
Over the decade tremendous progress have been made in Hearing aid (HA) and CI technologies, which have lead CI recipients to enjoy near-to-normal speech intelligibility in quiet conditions~\cite{zeng2008cochlear, natarajan2005auditory}. However, reduced speech intelligibility in the presence of background noise such as environmental sounds or competing talkers is a major complaint by CI recipients~\cite{friesen2001speech}; speech enhancement (SE) systems~\cite{loizou2007speech, wang2018speech, hansen2006speech, HansenOverview1999} can be used to reduce noise and provide better speech understanding for CI users.  

CI users receive spectral information with a limited number of encoding channels (normally 8 to 22 channels) which is far less than the frequency bands leveraged by normal hearing (NH) subjects \cite{Hansenembc2019}. This low-resolution representation of the speech signal increases the sensitivity of CI users to noisy conditions \cite{Mamun2019SID}. Therefore, speech enhancement algorithms have received growing attention in the CI research community. This study proposes a class of convolutional neural network (CNN)-based speech enhancement algorithms specifically designed for CI users.

Various algorithms have been developed by the researchers to address background noise for CI users. 
These SE algorithms can be broadly classified into two categories: (1) unsupervised, and (2) supervised. 

\textbf{Unsupervised} -- These approaches are based on estimating the statistical characteristics of both speech and noise signals. Spectral-subtraction~\cite{boll1979suppression}, Wiener filtering~\cite{khorram2012optimum}, and signal-subspace~\cite{ephraim1995signal} algorithms are well-known examples of this category. In the spectral-subtraction method~\cite{boll1979suppression}, clean speech spectrum is obtained by subtracting an estimation of the noise spectrum from the noisy spectrum. In the Wiener filtering method~\cite{almajai2011visually}, an optimal linear time-invariant filter that minimizes the mean-square-error (MSE) is estimated. In the signal-subspace method~\cite{ephraim1995signal}, the vector space of the noisy signal is decomposed into a signal-plus-noise subspace; SE is accomplished by removing the noise subspace. A major limitation of unsupervised methods is that they are based on specific noise or environment assumptions that are not necessary true for all domains and all noisy conditions.

\textbf{Supervised} -- These approaches consider the SE problem as a supervised machine learning task in which we train a data-driven function\footnote{In many cases, we train a component of the SE pipeline.} that takes noisy signal and estimates clean speech signal. In contrast to unsupervised techniques, these approaches require a training dataset that provides adequate knowledge about the distributions of speech and noise. Neural networks are general function approximators that have been widely used for supervise SE. Neural network-based SE (NNSE)~\cite{goehring2017speech} and long-term short-term memory recurrent neural network (LSTM-RNN)-based SE~\cite{sun2017multiple} are two successful examples of supervised SE.

NNSE algorithm, introduced in~\cite{goehring2017speech}, approximates frequency channels with higher SNR; reduces noise-dominated components of the signal from those channels and thus improves speech intelligibility for CI users. DNN-based SE algorithm, proposed in~\cite{xu2015regression}, uses log-power spectral features from speech data. The algorithm successfully reduces the non-stationary noise components without generating considerable musical artifacts. On the other hand, LSTM-RNN-based SE exploits a stack of LSTM layers to provide a direct mapping from noisy speech features to clean speech features~\cite{sun2017multiple}. This system has been reported to be more effective than DNN-based regression technique in modeling long-term acoustic context specifically in low SNR conditions. Although, existing supervised SE methods have achieved considerable success, these algorithms provide limited benefits for CI users due toseveral issues: first, standard DNN-based SE algorithms are not efficient in characterizing local temporal-spectral structures of speech signals; second, these algorithms have not been designed to effectively leverage feature space for CI users. Finally, they are only as effective as the diversity of speech, speaker, and noisy data available for model training. 

To deal with the first issue, this study proposes a set of CNN-based SE algorithms to improve the speech representation before the CI encoding processor. CNN is originally designed to consider local patterns of input signals by using a set of local connections \cite{fu2016snr}. CNN has been reported to be more effective than standard feed-forward neural network in many speech processing applications including speech emotion recognition~\cite{khorram2019jointly}, speech enhancement~\cite{fu2017raw} speech separation~\cite{Midea2019Probabilistic} and speaker recognition~\cite{bahmaninezhad2018compensation}. It is because CNN is able to deal with local temporal-spectral structures of speech and it can effectively separate the speech and the noise components of the noisy signals. 

We resolve the second issue by defining our feature space within the CI auditory space. We explore three different CNN-based schemes using knowledge of traditional SE methods. The vanilla CNN SE algorithm directly extracts features of the speech signal from the noisy speech; spectral-subtraction-style CNN (SS-CNN) SE algorithm estimates the noise from the input signal, subtracts it from the noisy signal and obtains the clean signal spectrum; Wiener-style CNN (Wiener-CNN) SE algorithm generates a weighting mask first, applies the mask to the incoming noisy signal and then generates the clean speech signal. This study also explores the causal structures of the proposed CNN-based algorithms. The causal CNNs use only previous samples of the signals and offer a viable option for real-time implementations of CCi-Mobile research platforms~\cite{Hansenembc2019}.

The paper is organized as follows. Sec. 2 explains the methodology of the proposed algorithms. It also includes the feature extraction technique leveraged in the proposed algorithms. Experimental setup and results of the proposed CNN-based algorithms are described in Sec. 3. We compare results of the proposed algorithms with results of the traditional methods. Finally, Sec. 4 concludes this study.

\section{Methodology}
In this section, we first briefly introduce the CI pipeline. We then explain details of the proposed SE algorithms. We also describe the computation of the objective speech intelligibility score designed for the CI users. We finally discuss existing baseline SE systems as well as different components of the proposed algorithms.   
\subsection{Cochlear Implant Signal Processing Pipeline}\label{sec:CI_pipeline}

The CI signal processing pipeline, used in our experiments contain the following steps: (1) Speech sampled at $16\ KHz$ is pre-emphasized to emphasize high frequency components of the signal. (2) A Hamming window of length $10\ ms$ with an overlap of $8.75\ ms$ is applied to the pre-emphasized signal. (3) A fast Fourier transform (FFT) of the extracted frames are used to calculate the spectral energy of each frequency bin. (4) The spectral energy signals are processed through a 22-channel filter-bank, that emulates the CI auditory system. This step generates a sequence of 22-dimensional features; we refer to these as \emph{``CI auditory features'} throughout this paper. (5) An 'n-of-m' CI signal processing strategy selects the 8 most important auditory features employed with a CIS-Continuous Interleaved Sampling strategy. (6) Finally, biphasic pulses are generated from the selected features and sent to the UTDallas CCi-MOBILE research interface board through electrical stimulations~\cite{Hansenembc2019}. These electrical stimulations can be visualized using electrodograms. An electrodogram (as shown in Fig.~\ref{fig:Electrodogram}) is a two-dimensional representation of time and electrode number (i.e., related to frequency bin over the auditory space). 

\subsection{Proposed Speech Enhancement Algorithm}

We propose to perform speech enhancement in the \emph{CI auditory feature space}, because this space has been specifically designed for CI users. To do so, we extract a set of 22-dimensional CI auditory features using the first 4 steps of the CI pipeline (explained in Sec.~\ref{sec:CI_pipeline}). The extracted features are passed to our proposed CNN-based SE algorithms to perform noise/interference reduction in the CI auditory feature space. Three
different CNN architectures are introduced in this paper. All the proposed architectures take noisy CI auditory features and estimate the clean speech features. The estimated features can be directly used to stimulate intracochlear electrodes in a CIS manner. In this paper, we use an objective speech intelligibility metric, rigorously tested for CI users, to evaluate speech intelligibility level in the CI auditory feature space.

\def\factor{0.8}
\begin{figure}[t]
\centerline{\includegraphics[width=\factor\linewidth]{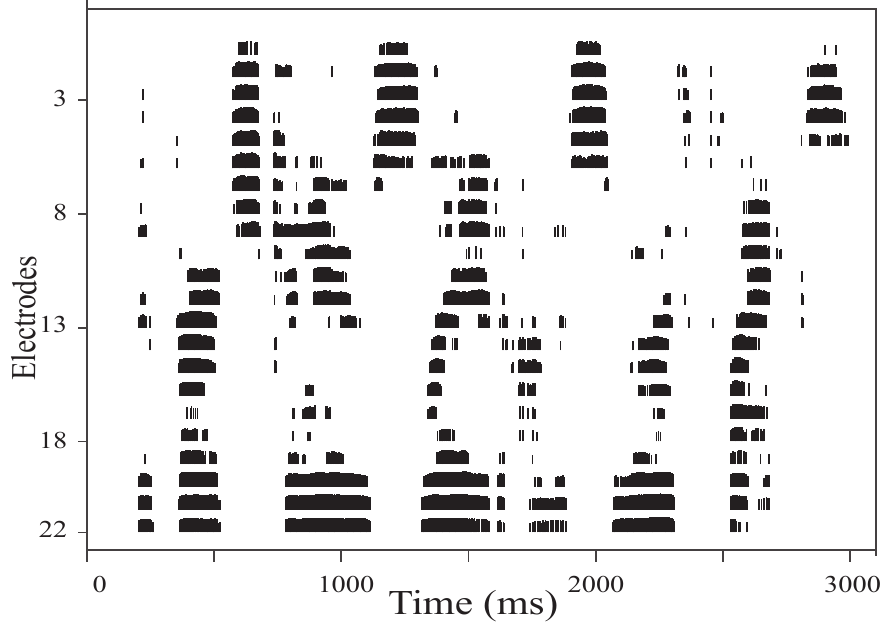}}
\caption{Cochlear implant electrode stimulation response shown as an electrodogram.
\vspace{-15pt}
}
\label{fig:Electrodogram}
\end{figure}

\subsubsection{Proposed CNN-based Architectures}
\begin{figure*}[t]
\centerline{\includegraphics[width=\linewidth]{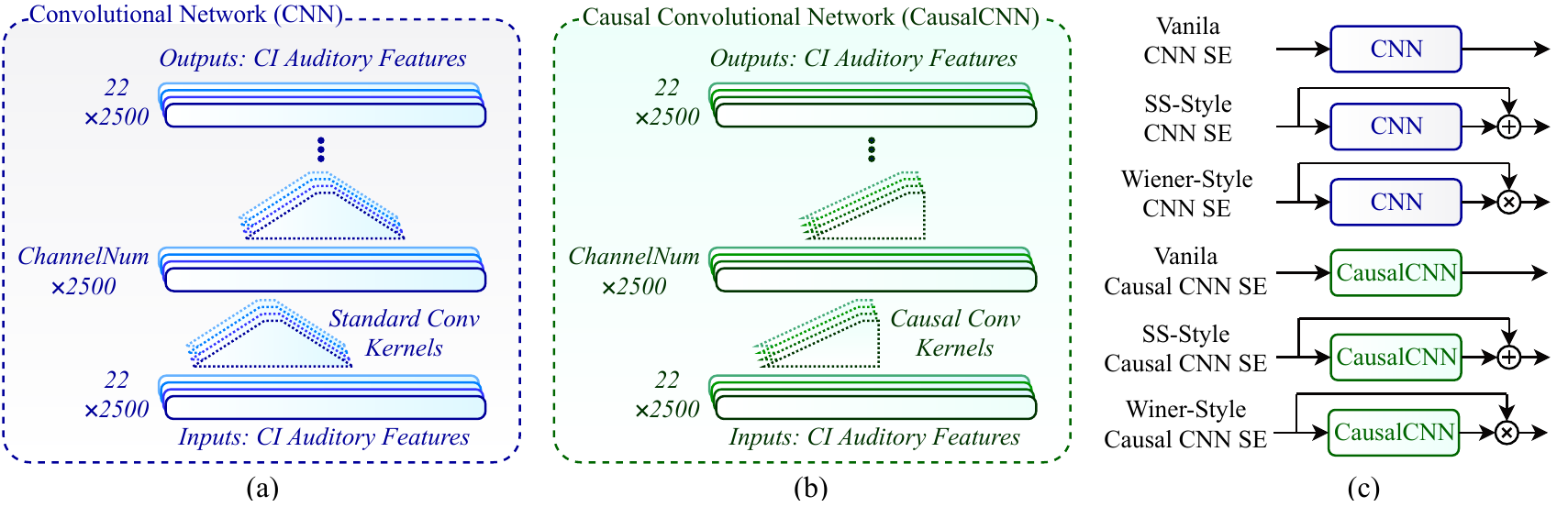}}
\caption{(a) Block diagram of the standard CNN used in this paper. (b) Block diagram of the causal convolutional network (Causal CNN) that leverages causal convolutional kernels in each layer. The causal kernels consider only previous samples of the signals. (c) Various SE systems proposed in this paper; we incorporate both CNN and causal CNN in three different network architectures: Vanila, spectral-subtraction-style, and Wiener-style architectures. We train all the networks in an end-to-end manner.}
\vspace{-15pt}
    \label{fig:methods}
\end{figure*}

In this study, we propose three different network architectures: Vanilla, SS-CNN and Wiener-CNN. We implement both the standard and the causal CNNs for the proposed architectures. Figure~\ref{fig:methods} shows the details of the proposed SE systems. 

\vspace{5pt}
\textbf{Vanilla CNN} --
This system directly estimates the clean speech features from the input noisy features. In this study, the input to the convolutional layer is a sequence of 22-dimensional CI auditory features. The CNN architecture consists of several convolution layers; each layer applies a set of linear finite impulse response (FIR) filters (kernels) to extract intermediate features; it also applies an activation function that enables the network to provide sophisticated, non-linear functionality. Parameters of the filters are trained during the training phase of the system using a gradient-based optimization algorithm.


\vspace{5pt}
\textbf{Spectral-subtraction-style CNN} --
Spectral subtraction is a simple yet effective method for SE \cite{boll1979suppression}. This method estimates the clean signal spectrum by subtracting estimated noise from the noisy signal spectrum. Traditional spectral subtraction methods are able to handle only stationary noises; however, in most real-life environments, background noise is non-stationary. To deal with this issue, in this paper, we propose a variation of spectral subtraction that uses CNNs to estimate the noise components. CNNs are able to extract noise components from the noisy signals which does not assume that the background noise is stationary. Fig.~\ref{fig:methods} highlights the difference between the Vanilla and the SS-Style systems. In the Vanilla system, the CNN directly estimates the speech components, yet in the the SS-Style system, the CNN estimates the noise components from the noisy part of the signal. 


\textbf{Wiener-style CNN} --
Conventional Wiener filter-based SE algorithm calculates the priori SNR based on the noise spectrum from the silent regions~\cite{almajai2011visually, khorram2012optimum}. It estimates an optimal filter (an optimal mask in the frequency domain) to generate the power spectrum of the clean speech. Similarly, the proposed system estimates an optimal mask using a CNN. The de-noised output features are then obtained by multiplying the incoming noisy features with the extracted mask.    

\textbf{Causal Convolutional Network} --
Conventional CNN based solutions use both past and future inputs to generate its output. Therefore, to implement a solution for real-time applications, we must store a buffer window of the input and then produce the output result (the size of the window must be at least half the size of the CNN receptive field~\cite{khorram2017capturing}. This process delay, which is a limiting factor for CI-related applications. To deal with this limitation, we also use a causal variation of the CNN (causal CNN) in our proposed architectures. In a causal CNN, each neuron for any particular time does not receive information from future neurons (Fig.~\ref{fig:methods}(b)). Therefore, causal CNN estimates the current features based on information obtained only from previous time frames and does not introduce a considerable delay o the output signal.

\subsection{Similarity Measure}

Several objective intelligibility metrics have been proposed in last few decades to measure speech intelligibility for hearing aids HA \cite{mamun2015prediction, akter2019predicting, mamun2018measuring} and CI \cite{yousefian2012predicting} users. This study uses the Envelope-based correlation measure (ECM) to calculate the objective intelligibility score for CI users. The ECM metric is a modulation-based index proposed for measuring speech intelligibility for subjects with cochlear implants \cite{yousefian2012predicting}. The metric uses the low frequency envelope modulation to predict the intelligibility score. Inputs to the ECM metric are the CI auditory features, and the output is the estimated speech intelligibility score which is a number between 0 and 1 ($ECM = 1$ indicates the highest intelligibility). It has been shown that ECM has the high correlation with subjective intelligibility scores~\cite{yousefian2012predicting}.

    
    


\subsection{Baseline Algorithms}
We compare the proposed methods with three conventional SE algorithms: (1) Minimum mean-square error log-spectrum amplitude estimator (Log-MMSE)~\cite{ephraim1985speech}, (2) Wiener filtering based on wavelet-threshold (Wiener-wt) \cite{hu2004speech}, and (3) Wiener filtering based on a priori SNR estimation  (Wiener-as) \cite{scalart1996speech}. The Log-MMSE algorithm captures the short-time spectral amplitude of the speech and utilizes MMSE to enhance the noisy signal. On the other hand, Wiener-as address the musical noise using wavelet thresholding and Wiener-as algorithm estimates the priori SNR to enhance the noisy speech.


\section{Experiments}
In this section, we compare the performance of the proposed and the baseline SE algorithms.

\subsection{Dataset}
We use \emph{``UT-Drive''} corpora to perform the experiments in this study~\cite{krishnamurthy2012vehicle}. UT-Drive is a large-scale database of noise signals collected across different vehicle platforms under a wide range of field driving conditions. The database contains two different sessions: off-road and on-road. In the off-road session, speech signals from the TIMIT database \cite{zue1990speech} were combined with the noise generated from different parts of a car including engine, wiper blade, right turn, left turn, AC, honk, etc. In the on-road session, a similar set of noise signals were collected, but when the car is driven with different speeds (i.e., 40, 50, and 60 mph).


\subsection{Experimental Setup}
In our experiments, TIMIT signals are distorted with a part of the UT-Drive database containing noise signals of a Mitsubishi Galant (2002) and a Nissan Sentra (2008) cars. For both cars, AC is off and all windows are closed during the recording process. We refer to the Mitsubishi Galant and the Nissan Sentra as ``Car 1'' and ``Car 2'', respectively, throughout this paper. The TIMIT database includes 6300 utterances from different speakers. We split this dataset into three non-overlapping sets: train, development, and test sets. Train set includes 3150 utterance and is used to train our CNNs. Development set contains 1575 utterances and is used to tune the network hyper-parameters. Test set contains 1575 utterances and is used to compare the performance of different systems.  

We build our CNNs using Keras with a Tensor-Flow back-end. We train all CNNs by optimizing the mean square error (MSE) metric through the Adam optimizer~\cite{gideon2017progressive}. Each network is trained for 300 epochs and the best epoch is selected based on the development MSE. Parameter tuning procedure results in a CNN architecture with 7 convolutional layers (6 hidden layers and 1 output layer). Each convolutional layer contains the same number of neurons with 65 kernels. Finally, 'tanh' and 'linear' activation functions are used for the intermediate and output layers, respectively. 

\def\factor{0.9}
\begin{figure}[t]
\centerline{\includegraphics[width=\factor\linewidth]{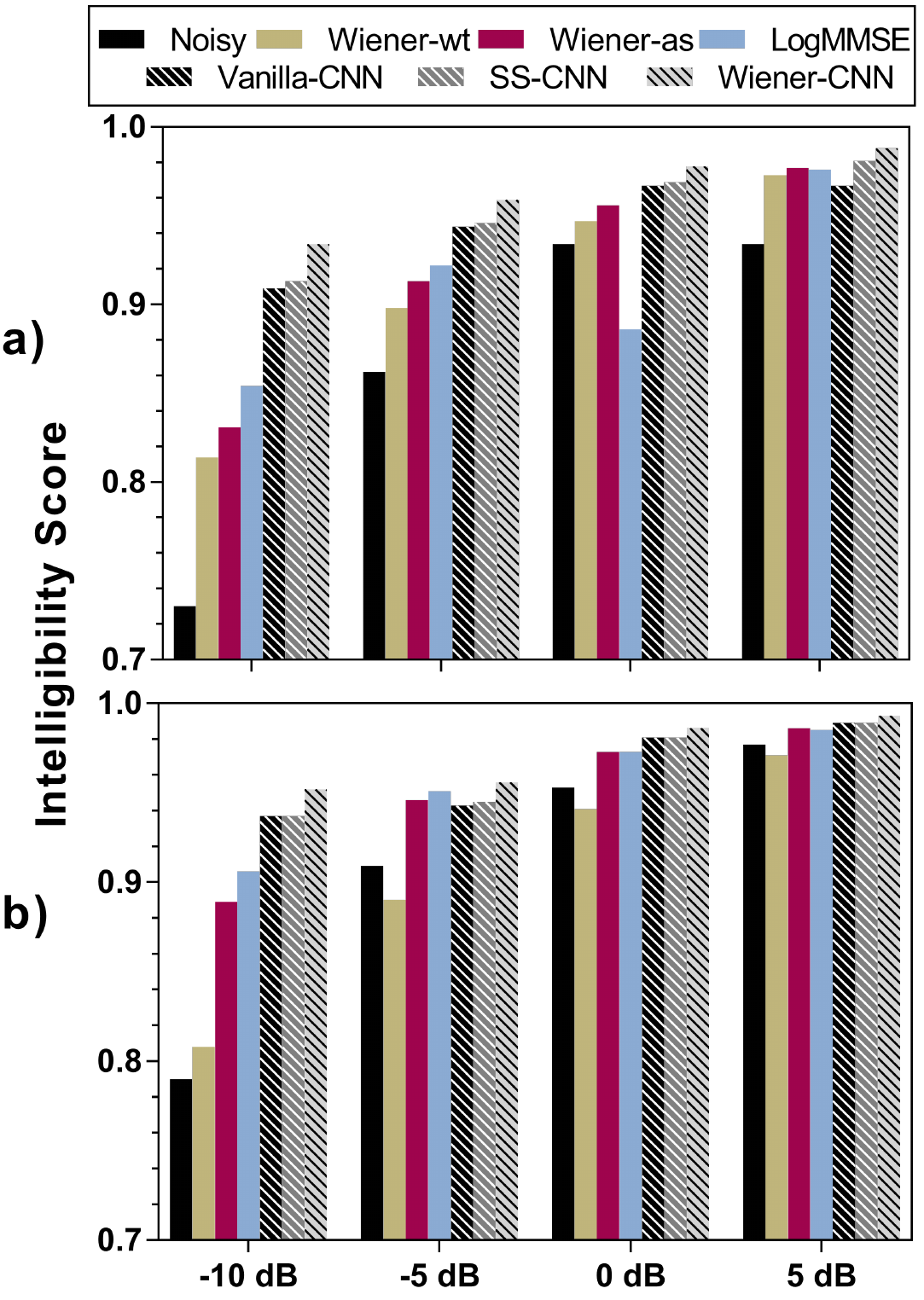}}
\caption{Mean speech Intelligibility score based on the ECM measure as a function of SNR for proposed Non-causal SE algorithms. Noise environments: (a) Car 1: Mitsubishi Galant (2002) (b) Car 2: Nissan-Sentra (2008).\vspace{-15pt}}

\label{fig:NonCausal}
\end{figure}

\subsection{Results}
In this section, we use ECM to evaluate the performance of the proposed systems. We consider four SNRs in our experiments: -10dB, -5dB, 0dB and 5dB. We note here that the majority of car noise resides at low frequencies, so SNRs of -10dB and -5dB still have some intelligible speech content. For each SNR value, we artificially add two types of noises explained in the previous section (referred as Car1 and Car2 noises in our experiments). Noisy speech signals are enhanced with the proposed and the baseline SE algorithms. Figures~\ref{fig:NonCausal} and \ref{fig:Causal} summarize the ECM results of our experiments.  

Fig.~\ref{fig:NonCausal} shows the performance of six different algorithms (three non-causal versions of the proposed CNN and three baseline systems) in two environments (Car 1 and Car 2). In general, all algorithms exhibit similar performance (similar ECM intelligibility score) for the high SNR value of 5dB; however, the proposed algorithms perform better than the baseline systems for the lower SNR values (-10dB and -5dB). Moreover, Wiener-CNN-based SE outperform other proposed systems
and therefore it is suggested that the Wiener-style architecture be used for CI users.



We also exploit causal versions of the proposed algorithm to enhance the noisy speech signals. Fig.~\ref{fig:Causal} shows results with causal CNN architectures. Results show that causal versions of the proposed algorithms are able to significantly improve speech intelligibility in CI features domain. Moreover, the proposed Wiener-CNN SE (causal) algorithm outperform other causal CNNs and baseline systems. Therefore, we expect that the proposed Wiener-CNN SE algorithm to be a potential option to enhance speech intelligibility for CI users in naturalistic environments.


\def\factor{0.9}
\begin{figure}[t!]
\centerline{\includegraphics[width=\factor\linewidth]{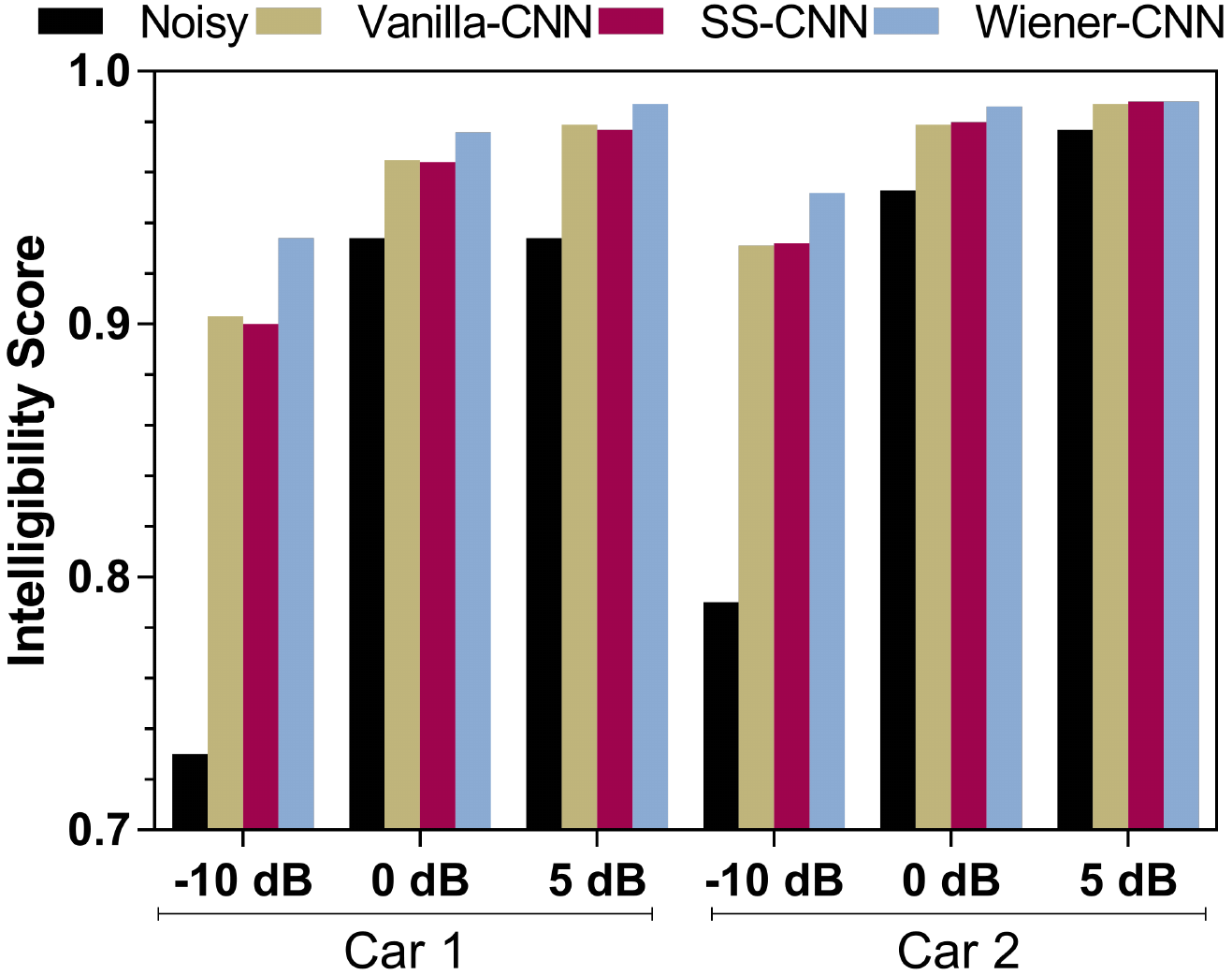}}
\caption{
Denoising performance of the causal versions of the proposed algorithms in Car 1 and Car 2 noise environments.}
\vspace{-15pt}
\label{fig:Causal}
\end{figure}
\vspace{-10pt}

\section{Conclusion}
The main goal of this study has been to propose a set of CNN-based SE algorithms that could be useful for CI users in naturalistic noisy conditions. The contribution of this study is threefold.  First, we extracted speech features from noisy signal based on CI auditory features. The extracted features were used in the proposed SE algorithms. Second, we proposed a set of novel speech enhancement algorithms based on CNN (both non-causal and causal versions), which were designed to estimate clean auditory features from the noisy speech for CI users. The reconstructed features were used to generate electrodograms that reflect the actual CI auditory stimulation. Third, our experimental results show that non-causal CNN-based SE algorithms provides better speech enhancement compared to causal versions. The results also show that the highest ECM score is obtained for the non-causal Wiener-CNN algorithm. In this paper, we investigated three different CNN-based architectures for SE in CI auditory feature space; however, other neural network architectures such as recurrent networks can also be applied and may provide better performance for some noise types. In our future work, we plan to explore other network architectures in CI auditory feature space.

\vspace{-10pt}
\section{Acknowledgement}
This work was primarily supported by a National Institute on Deafness and Other Communication Disorders (NIDCD) Grant (No. R01 DC016839-02).

\bibliographystyle{IEEEtran}

\bibliography{mybib}


\end{document}